# Exploring the Multi-modal Demand Dynamics During Transport System Disruptions

A. Ali Shateri Benam[a],[*1], B. Angelo Furno[a], C. Nour-Eddin El Faouzi[a]

[a] *LICIT-ECO7, ENTPE, Univ. Gustave Eiffel*

## Abstract

Various forms of disruption in transport systems perturb urban mobility in different ways. Passengers respond heterogeneously to such disruptive events based on numerous factors. This study takes a data-driven approach to explore multi-modal demand dynamics under disruptions. We first develop a methodology to automatically detect anomalous instances through historical hourly travel demand data. Then we apply clustering to these anomalous hours to distinguish various forms of multi-modal demand dynamics occurring during disruptions. Our study provides a straightforward tool for categorising various passenger responses to disruptive events in terms of mode choice and paves the way for predictive analyses on estimating the scope of modal shift under distinct disruption scenarios.

## Introduction

Human mobility in the urban context follows specific patterns and cycles [1]. These patterns are prone to short-term divergences or long-term changes. Disruptive events such as extreme weather conditions, road or public transport accidents, or large-scale events can disrupt the usual flow of urban travel. In response to such events, passengers exhibit heterogeneous responses depending on various factors, including their original mode of choice and the nature and timing of the disruption [2]. While these responses may include the cancellation or delay of trips, this study seeks to investigate the inter-modal shifts that occur during such disruptive events.

Previous research has examined the passenger responses to disruptions in transport systems using surveys [3,4] and mobile or public transport ticketing data to analyse the demand dynamics between two specific modes during certain events [5,6]. However, current literature lacks an integrated, automated approach that goes beyond single-kind disruption events and accounts for multiple urban transport modes. This study aims to follow that approach in identifying disruptions and exploring the associated multi- and inter-modal demand dynamics.

Utilising multi-source data from various entities, we have developed methods for detecting and classifying anomalous multi-modal travel demand using machine learning techniques. By leveraging urban mobility data related to the city of Lyon, France, we have identified several forms of passenger response to disruptions through travel demand anomalies. Our approach is innovative in its data-driven analysis of complex multi-modal demand dynamics during disruptive events. This study aims to inform the development of more effective strategies for mitigating the negative impacts of transport network disruptions.

## Data & Methodology

We present a methodological framework for exploring the multi-modal dynamics through anomalous demand detection. We use road traffic, public transport and shared-bike data at our disposal. Our road data comes from a Floating Car Dataset (FCD), which provides timestamped geo-localised data collected via GPS from moving vehicles. Each observation in the FCD data captures information such as the corresponding vehicle ID's speed, direction, and map-matched link identifier with an average observation frequency of 30 seconds. The public transport data is obtained from Lyon's public transport provider and includes ticket validation data for trams, metro, and bus lines. Each validation record in the dataset includes temporal information of said validation with its corresponding line. Finally, we incorporate the shared-bike data from Lyon's Velo'V bike-sharing service, which contains information about bike retractions and returns and their stations. We follow our methodology through 4 steps, shown in Figure 1.

---

([1]) ali.shateribenam@entpe.fr

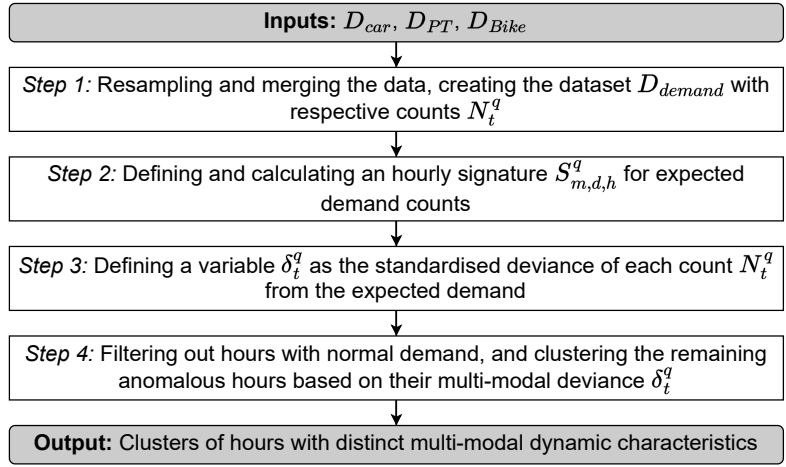

Figure 1. Methodology flowchart

**Step 1:** First, we constructed a compressed representation of the typical demand for each transport mode by unifying our multi-source datasets. By aggregating observations hourly, we combined three datasets ($D_{car}$, $D_{PT}$, and $D_{Bike}$) into one unified dataset, $D_{demand}$. The hourly number of observations in the FCD dataset is related to the city-wide number of moving cars on the road network and is thus assimilated to an hourly count for car demand. The number of hourly ticketing validations for public transport modes is associated with the hourly demand for the respective mode. The hourly number of shared-bike retractions is used to indicate shared-bike demand. The final pre-processed dataset is denoted as $D_{demand} = \{N_t^q\}$. $N_t^q$ represents the observed demand count for mode $q$ on the generic 1-hour time slot $t$ from the data availability period, and $q \in Q$ = {bus, tram, metro, bike, car}.

**Step 2:** We adapt and extend the approach initially proposed in [7] by defining a rolling signature describing the expected demand on each mode for a given reference period (one week). Such a signature, defined on 24*7 hourly time period, is a compressed representation computed from previous and future observations of demand counts. Specifically, the expected behaviour for a given week $m$, is computed over a temporal support ($W_m^K$, $d$, $h$). $W_m^K \subset M$ represents the subset of weeks, of size K (a hyper-parameter of our approach), from the available set of weeks (M), related to a specific temporal reference point (a given week number $m$), with the reference point itself excluded from the subset of weeks. $d \in \Delta$, represents the weekday variable from an ordered set of selected weekdays $\Delta$. $h \in H$, represents the hour of the day variable from an ordered set of selected hours $H$. It is worth noting that the ($W_m^K$, $d$, $h$) support identifies a set of time slots that will be considered for aggregation during the computation of the signature. Based on the temporal support ($W_m^K$, $d$, $h$) and transport mode $\in Q$, the signature element $S_{(m,d,h)}^q$ is defined as Equation 1.

$$S_{(m,d,h)}^q = \mu_{(W_m^K,d,h)}^q \pm \lambda_{(W_m^K,d,h)}^q \qquad \text{Equation 1}$$

Setting the number of weeks K for our temporal support to 4, $\mu_{(W_m^4,d,h)}^q$ represents the average demand counts for the mode $q$, hour $h$, and weeks $m-2$, $m-1$, $m+1$, and $m+2$. The standard deviation for the same set of hours is denoted as $\sigma_{(W_m^K,d,h)}^q$, and the range amplitude of the signature $\lambda_{(W_m^K,d,h)}^q$ is defined as the product of a coefficient $\alpha$ and the hourly standard deviation $\sigma_{(W_m^K,d,h)}^q$. Through a sensitivity analysis based on the knee curve method [8], we determined the standard deviation hyper-parameter $\alpha$ for the range amplitude to be 4, therefore $\lambda_{(W_m^K,d,h)}^q = 4 \times \sigma_{(W_m^K,d,h)}^q$. Around 15% of the hours in 2019 and 2020 fall outside the expected signature produced by this range amplitude.

**Step 3:** To assess the distance of observed demand from the expected average demand for a particular mode at a given hour, a continuous variable $\delta_t^q$ is defined as formulated in Equation 2. By pivoting the data on the deviance variable $\delta_t^q$, we produce a dataset $D_{deviance}$ with transport modes for columns, hours for rows, and standardised deviance of expected demand for values.

$$\delta_t^q = \frac{N_t^q - \mu_{(W_m^K,d,h)}^q}{\sigma_{(W_m^K,d,h)}^q} \qquad \text{Equation 2}$$

**Step 4:** As we are solely looking through anomalous hours for their categorisation, we omit hours in which demand in all modes falls inside the expected signature in a new dataset denoted as $D_{anomaly}$. This dataset is produced by filtering out regular hours; thus, it only includes hours where at least one mode's absolute value of deviance $|\delta_t^q|$ is bigger than the range amplitude of the signature $\lambda_{(W_m^K, d, h)}^q$. We then apply clustering to this dataset to categorise various anomaly forms regarding multi-modal demand dynamics.

## Results

Following our previous study, we hypothesised to observe distinct types of multi-modal dynamics in disruptions, and we applied agglomerative clustering with a cosine distance to our $D_{anomaly}$ dataset. As our research focuses on discrete forms of multi-modal passenger response, we adopt the Davies-Bouldin method for obtaining an optimal number of clusters [9]. The method indicates 11 as the optimal number for clusters to be as non-identical as possible.

We conducted clustering and labelled each hour with the cluster number. We selected all hours belonging to each cluster to exhibit how each cluster manifested a particular multi-modal dynamic. We averaged the deviance $\delta_t^q$ for each mode and produced a demand dynamic profile for each cluster. We then used radar plots to illustrate these multi-modal dynamic profiles. Additionally, we labelled the dates and the hours of anomalies belonging to each cluster to explore their temporal correspondence to traceable events.

The multi-modal dynamics of clusters were adequately explainable by ground truth (news and weather data). We present three clusters of anomalous data and their interpretation (other clusters are not reported due to space limitations). The radar plot for each cluster displays different modes' deviance $\delta_t^q$ averaged among the hours belonging to the corresponding cluster. For example, a value of +1 on the plot indicates that, on average, demand counts belonging to the cluster's hours have been one standard deviation above the temporal support's average demand.

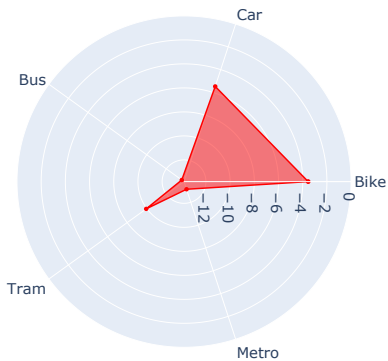 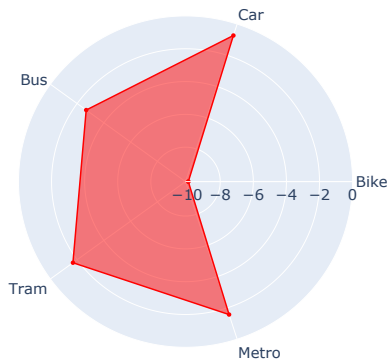 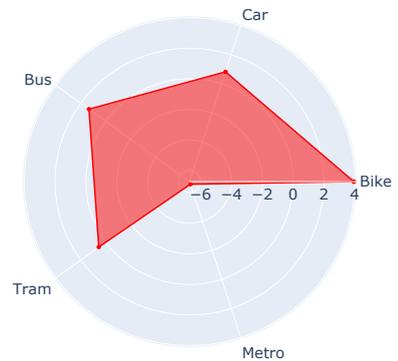

*Figure 2. Average multi-modal demand deviance in Cluster 1*  *Figure 3. Average multi-modal demand deviance in Cluster 2*  *Figure 4. Average multi-modal demand deviance in Cluster 3*

The radar plot representing the average deviance of each mode in *Cluster 1* is exhibited in Figure 2. The plot indicates negative deviance from the expected average demand in all modes. However, the demand for public transport modes is much lower than the expected average demand. *Cluster 1* covered around 36% of all anomalous hours through 2019 and 2020. The hours of *Cluster 1* were not scattered and were mostly continuous for several hours. Using France's holiday calendar for 2019 and 2020, we verified that the hours belonging to this cluster were mainly affiliated with national holidays, such as Christmas, New Year's, and Easter holidays.

Figure 3 illustrates the average deviance of each mode in *Cluster 2*, which covers 17.5% of all anomalous hours through 2019 and 2020. This plot shows the demand dynamics of anomalous hours in which a slight decrease in demand for all modes and a sharp decrease in shared-bike use are observed. Building from our previous study, we hypothesised that this cluster might represent instances of anomaly where the use of shared bikes has been negatively affected by precipitation. We made use of Lyon's meteorological data to test this hypothesis. The meteorological data at our disposal consists of hourly recorded information on precipitation, wind and their properties. To put our hypothesis to the test, we employed a one-sample student's t-test (a reference here) to determine if the precipitation rate is significantly higher during the hours marked with *Cluster 2's* label. The t-test results prove that the precipitation during Cluster 2 hours is significantly higher than the entire hours, with a t-value of 5.03 and a P-value of $7.29*10^{-7}$.

*Cluster* 3 covers around 1% of all anomalous hours, and as shown in Figure 4, the hours in this cluster exhibit metro validations with an average negative deviance of -6. At the same time, other public transport modes and road traffic exhibit slight positive average deviance, while shared bikes demonstrate significant positive deviance. Such dynamics may reflect disruptions in the metro service, leading passengers to shift to other modes. Our hypothesis was verified by investigating the news articles through calendar hours labelled with Cluster 3. These hours corresponded to metro closure events where some metro lines or stations were closed (such as terrorist attacks, bomb suspicions, and human accidents). Another point in the calendar belonging to this cluster was Lyon's festival of lights, where metro service was announced free for the afternoon, resulting in zero metro validations.

## Conclusion

This study explores the complex network-wide multi-modal demand dynamics under disruptions. Specifically, we examined the case of Lyon by unifying multi-source data and defining a normal/expected range of hourly demand for each mode of transportation. By identifying anomalous instances where demand fell outside this range, we applied clustering techniques to investigate the various distinct multi-modal passenger responses to these disruptions. The clusters formed from the abovementioned method revealed significant variations in their multi-modal dynamics. These dynamics were analysed on ground truth, three examples were presented, and their interpretations were explained. We plan to continue this study by diving more deeply into a more predictive approach by training machine-learning regression models over larger instances of historical multi-modal demand datasets. A decision-maker could leverage the latter to determine, in advance, the expected travel demand increase/decrease of one mode when one or multiple other modes are expected to be disrupted.